%% file: main.tex
\documentclass[conference]{IEEEtran}
\usepackage[noadjust]{cite}
\usepackage{graphicx,color}
\usepackage[dvipsnames]{xcolor}
\usepackage{amsmath,amsbsy,amsfonts,amssymb}
\usepackage{mathrsfs}
\usepackage{mathtools}
\mathtoolsset{showonlyrefs}
\ifCLASSOPTIONcompsoc
\usepackage[caption=false,font=normalsize,labelfont=sf,textfont=sf]{subfig}
\else
\usepackage[caption=false,font=footnotesize]{subfig}
\fi

\IEEEoverridecommandlockouts 

\usepackage{lipsum}  
\usepackage{dblfloatfix}

\usepackage{etoolbox}
\usepackage{tikz}
\newrobustcmd*{\squareA}[1]{\tikz{\filldraw[draw=#1,fill=#1] (0,-0)
rectangle (0.1cm,0.14cm);}}
\newrobustcmd*{\mycircle}[1]{\tikz{\filldraw[draw=#1,fill=#1] (0,-0.3) circle [radius=0.08cm];}}

\newcommand{\figr}{Fig.~}
\newcommand{\secr}{Sec.~}

\usepackage{acro}
\input{acronyms.tex}
\input{./Figures/colour_map_AM.tex}

\begin{document}

\title{\huge Age of Incorrect Information in Random Access Channels without Feedback}
\author{
\IEEEauthorblockN{Andrea Munari\\
\IEEEauthorblockA{Institute of Communications and Navigation, German Aerospace Center (DLR), Wessling, Germany\\
}}
}
\date{}
\maketitle
\thispagestyle{empty}
\pagestyle{empty}

\begin{abstract}
We focus on a system in which a set of two-state Markov sources report status update to a common receiver over a shared wireless channel. Inspired by practical IoT networks, we consider three variations of ALOHA as medium access protocol: i) a random approach in which a source transmits regardless of its status, ii) a reactive scheme in which updates are sent only when a source changes state, and iii) a hybrid solution which blends the two possibilities. We consider different criteria to capture the ability of the receiver to maintain an accurate perception of the tracked processes: average age of incorrect information (AoII), probability of missed detection (i.e., of not detecting a source transition), and average duration of intervals over which the receiver lingers with erroneous knowledge. We provide closed form analytical expressions for all the metrics, highlighting non-trivial trade-offs and providing useful protocol design hints. 
\end{abstract}

\input{notation.tex}

\input{introduction.tex}
\input{sysModel.tex}
\input{analysis.tex}

\input{results.tex}
\input{conclusions.tex}

\bibliographystyle{IEEEtran}
\bibliography{IEEEabrv,biblio_RandomAccess,biblio_AoI}

\end{document}

%% file: acronyms.tex
\DeclareAcronym{AWGN}{short = AWGN ,long = additive white gaussian noise}
\DeclareAcronym{AoI}{short = AoI ,long = age of information}
\DeclareAcronym{AoII}{short = AoII ,long = age of incorrect information}

\DeclareAcronym{CDF}{short = CDF ,long = cumulative distribution function}
\DeclareAcronym{CRA}{short = CRA ,long = contention resolution ALOHA}
\DeclareAcronym{CRDSA}{short = CRDSA ,long = contention resolution diversity slotted ALOHA}
\DeclareAcronym{CSA}{short = CSA ,long = coded slotted ALOHA}
\DeclareAcronym{C-RAN}{short = C-RAN ,long = cloud radio access network}
\DeclareAcronym{DAMA}{short = DAMA ,long = demand assigned multiple access}
\DeclareAcronym{DSA}{short = DSA ,long = diversity slotted ALOHA}
\DeclareAcronym{eMBB}{short = eMBB ,long = enhanced mobile broadband}
\DeclareAcronym{FEC}{short = FEC ,long = forward error correction}
\DeclareAcronym{GEO}{short = GEO ,long = geostationary orbit}
\DeclareAcronym{GF}{short = GF ,long = generating function}
\DeclareAcronym{IC}{short = IC ,long = interference cancellation}
\DeclareAcronym{IoT}{short = IoT ,long = Internet of things}
\DeclareAcronym{IRSA}{short = IRSA ,long = irregular repetition slotted ALOHA}
\DeclareAcronym{LEO}{short = LEO ,long = low Earth orbit}
\DeclareAcronym{MAC}{short = MAC ,long = medium access}
\DeclareAcronym{mMTC}{short = mMTC ,long = massive machine-type communications}
\DeclareAcronym{MC}{short = MC ,long = Markov chain}
\DeclareAcronym{PDF}{short = PDF ,long = probability density function}
\DeclareAcronym{PER}{short = PER ,long = packet error rate}
\DeclareAcronym{PLR}{short = PLR ,long = packet loss rate}
\DeclareAcronym{PMF}{short = PMF ,long = probability mass function}
\DeclareAcronym{RA}{short = RA ,long = random access}
\DeclareAcronym{rv}{short = r.v. ,long = random variable}
\DeclareAcronym{SA}{short = SA , long = slotted ALOHA}
\DeclareAcronym{SIC}{short = SIC ,long = successive interference cancellation}
\DeclareAcronym{SNR}{short = SNR ,long = signal-to-noise ratio}
\DeclareAcronym{SFG}{short = SFG ,long = signal flow graph}
\DeclareAcronym{TDM}{short = TDM ,long = time division multiplexing}

%% file: Figures/colour_map_AM.tex
\tikzstyle{dashed} 	= [dash pattern=on 8pt off 5pt]
\tikzstyle{dashed_dense} 	= [dash pattern=on 5pt off 1.5pt]
\tikzstyle{dashddotted} 	= [dash pattern=on 11pt off 3pt on \the\pgflinewidth off 3pt on \the\pgflinewidth off 3pt on \the\pgflinewidth off 3pt]
\tikzstyle{dashdotted} 	= [dash pattern=on 7pt off 3pt on \the\pgflinewidth off 3pt]
\tikzstyle{dottedAM}	= [dash pattern=on 2pt off 4pt]

\definecolor{col_B}{HTML}{88CCEE}
\definecolor{col_C}{HTML}{44AA99}
\definecolor{col_D}{HTML}{117733}
\definecolor{col_E}{HTML}{999933}
\definecolor{col_F}{HTML}{DDCC77}
\definecolor{col_G}{HTML}{CC6677}
\definecolor{col_H}{HTML}{882255}
\definecolor{col_I}{HTML}{AA4499}
\definecolor{col_J}{RGB}{239,151,8}
\definecolor{col_K}{HTML}{56B4E9}
\definecolor{col_A}{RGB}{133,149,225}
\definecolor{col_W}{HTML}{A7C957}
\definecolor{col_Y}{HTML}{F4ACB7}
\definecolor{col_Z}{HTML}{FF8800}

\definecolor{oxygenblue}{rgb}{0.25, 0.41, 0.8}
\definecolor{apricot}{HTML}{FBB982}
\definecolor{redOrange}{HTML}{F26035}
\definecolor{darkGreen}{HTML}{009B55}
\definecolor{darkRed}{RGB}{142,6,59}
\definecolor{mildBlue}{RGB}{133,149,225}
\definecolor{strongerPink}{RGB}{211,63,106}
\definecolor{lightRed}{RGB}{187,119,132}
\definecolor{strongPink}{RGB}{224,123,145}
\definecolor{lightPink}{RGB}{247,156,212}

\definecolor{graphite}{HTML}{594D5B}
\definecolor{grey}{HTML}{6D616F}
\definecolor{silver}{HTML}{ADADC9}
\definecolor{pebble}{HTML}{333333}
\definecolor{coin}{HTML}{9897A9}
\definecolor{dove}{HTML}{7C6E7F}
\definecolor{rust}{HTML}{A4969F}
\definecolor{cloud}{HTML}{C5C6D0}

%% file: notation.tex
\newtheorem{prop}{Proposition}
\newtheorem{lemma}{Lemma}
\DeclareRobustCommand{\stirling}{\genfrac\{\}{0pt}{}}

\newcommand{\pr}{\ensuremath{\mathbb P}}
\newcommand{\expOp}{\ensuremath{\mathbb E}}
\newcommand{\de}{\mathrm{d}}

\newcommand{\pmc}{\ensuremath{q}}
\newcommand{\pTrans}[2]{\ensuremath{q}_{#1#2}}

\newcommand{\mcn}{\ensuremath{X_n}}
\newcommand{\est}{\ensuremath{\hat{X}}}
\newcommand{\estn}{\ensuremath{\hat{X}_n}}
\newcommand{\err}{\ensuremath{\delta}}
\newcommand{\errn}{\ensuremath{\err_n}}
\newcommand{\treset}{\ensuremath{R}}
\newcommand{\trise}{\ensuremath{S}}
\newcommand{\errper}{\ensuremath{W}}
\newcommand{\errperval}{\ensuremath{w}}
\newcommand{\averr}{\ensuremath{\overbara{W}}}
\newcommand{\corrper}{\ensuremath{Y}}
\newcommand{\corrperval}{\ensuremath{y}}
\newcommand{\pmiss}{\ensuremath{\mathsf P_{\mathsf m}}}

\newcommand{\pchange}{\ensuremath{\alpha_c}}
\newcommand{\pstale}{\ensuremath{\alpha_s}}
\newcommand{\prand}{\ensuremath{\alpha}}
\newcommand{\avgAct}{\ensuremath{\rho}}
\newcommand{\psucc}{\ensuremath{\gamma}}

\newcommand{\qZZ}{\ensuremath{\pTrans{0}{0}}}
\newcommand{\qZO}{\ensuremath{\pTrans{0}{1}}}
\newcommand{\qOZ}{\ensuremath{\pTrans{1}{0}}}
\newcommand{\qOO}{\ensuremath{\pTrans{1}{1}}}
\newcommand{\statZ}{\ensuremath{\pi_0}}
\newcommand{\statO}{\ensuremath{\pi_1}}
\newcommand{\statZZ}{\ensuremath{\pi_{0,0}}}
\newcommand{\statOO}{\ensuremath{\pi_{1,1}}}
\newcommand{\statZO}{\ensuremath{\pi_{0,1}}}
\newcommand{\statOZ}{\ensuremath{\pi_{1,0}}}
\newcommand{\avgQ}{\ensuremath{\overbar{q}}}

\newcommand{\nodes}{\ensuremath{\mathsf M}}

\newcommand{\tru}{\ensuremath{\mathsf S}}
\newcommand{\pen}{\ensuremath{\Omega}}
\newcommand{\load}{\ensuremath{\mathsf G}}
\newcommand{\avPen}{\ensuremath{\bar{\pen}}}
\newcommand{\aoii}{\ensuremath{\Omega}}
\newcommand{\Aoii}{\ensuremath{\overbara{\Omega}}}

\newcommand{\tStamp}{\ensuremath{\tau}}
\newcommand{\overbar}[1]{\mkern 1.5mu\overline{\mkern-3mu#1\mkern0.5mu}\mkern 1.5mu}
\newcommand{\overbara}[1]{\mkern 1.5mu\overline{\mkern0.1mu#1\mkern-0.1mu}\mkern 1.5mu}

%% file: introduction.tex
\section{Introduction} \label{sec:intro}

Remote tracking of physical processes represents a key use case in \ac{IoT} systems, with applications ranging from environmental and industrial monitoring to smart city and agriculture. These scenarios are characterized by the deployment of a large number of wireless sensor nodes, which report the status of a quantity of interest towards a common collector (sink) for further processing. In this context, the ultimate goal of the network is to provide an accurate perception of the monitored sources at the receiving endpoint, in order to enable proper actuation or decision making. The task is in turn not trivial, as commercial solutions \cite{LoRa,SigFox} often resort to simple uncoordinated medium access policies based on variations of the ALOHA paradigm \cite{Abramson77:PacketBroadcasting}, which are prone to frequent packet losses due to interference and collisions.

Inspired by these settings, research efforts have recently led to the definition of novel performance indicators. 
A pioneering role in the field was played by the age of information (AoI) \cite{Kaul11_Globecom}, which captures the freshness of the last available piece of information at the sink. Thanks to a smart yet simple mathematical definition, AoI has been thoroughly studied, leading to the derivation of fundamental insights in a number of IoT setups \cite{Yates20_Survey,Uysal20_TIT,Shirin22_TIT}. On the other hand, the metric only deals with timeliness, and falls short in characterizing systems where the value of monitored processes is relevant. To overcome this limitation, recent works have focused on different evaluation criteria. such as value of information \cite{Soleymani20_valueInfo}, entropy-based indicators \cite{Cocco23:ITW,Liew22_TIT,Pappas22_WiOpt}, or \emph{age of incorrect information} (AoII) \cite{Ephremides19_AoII}. The latter, in particular, goes beyond AoI by considering a metric that grows with time when the estimate available at the collector is not accurate, and foreseeing no penalty otherwise. 

Since its introduction, AoII has received significant attention for both point-to-point and multi-access systems. Among many relevant works, \cite{Joshi21_WiOpt,Ephremides21_Globecom} propose threshold-based policies in single-link settings, considering unreliable channels and power constraints, whereas the role of hybrid ARQ is tackled in \cite{Ephremides23_arXiv}. Optimal scheduling algorithms to allocate resources among multiple terminals were derived in \cite{Kriouile21_ISIT}, minimizing AoII via tools for partially observable Markov decision processes. In a similar setting, \cite{Elif23_arXiv_AoII} studies the metric for both push- and pull-based system, introducing the notion of query AoII.

On the other hand, the study of IoT networks that rely on uncoordinated medium access under the lens of AoII remains to date largely unexplored, leaving an important gap towards a thorough understanding of the behavior of commonly employed protocols. A first important step in this direction was taken in \cite{Chiariotti23:ICC}, considering a slotted ALOHA contention in which terminals monitor a set of Markov processes. The work leverages the availability of feedback, i.e., the possibility for a node to be aware of the current AoII  at the receiver, and tunes the channel access probability in a dynamic way to favor transmissions from sources with high penalty values. Dealing with a complex optimization problem, the proposed solution resorts to some heuristics, refining a threshold-based policy to significantly improve AoII performance. 

The present paper approaches the setting from a different angle, observing that a slot-wise feedback is often not implemented in practical IoT systems, entailing a cost in terms of energy expenditure and complexity that may not be viable for low-power, low-complexity sensor nodes \cite{Munari22_Globecom}. Starting from this remark, we focus on a setup in which a set of independent two-state Markov sources are monitored by terminals that send status updates to a common receiver over a slotted ALOHA channel without feedback. For such configuration, we tackle three variations of the access policy, studying a solution that decouples the access probability from the state of a source, a reactive approach which foresees a node to transmit only when a change in the source state takes place, and a hybrid scheme that blends the two possibilities. In all cases we analytically compute closed form expressions for the average AoII for both symmetric and asymmetric Markov sources, which allow to derive the optimal access probabilities. The study pinpoints some non-trivial take-aways, clarifying how a minimization of the AoII may come at the expense of a throughput reduction for the ALOHA channel. In addition, we evaluate the performance of the system in terms of \emph{missed detection probability} and average duration of a period over which the receiver has an erroneous estimate of a source state. We argue that such metrics are relevant in many practical use cases, e.g., when actuation or alarms shall be triggered, and we show how their optimization can require protocol configurations that depart from those minimizing AoII. 

\emph{Notation:} We denote a discrete r.v. and its realization by capital and lower-case letters, respectively, e.g. $X$, $x$. The probability mass function (PMF) of the r.v. is indicated by $p(x)$. For finite state Markov chains, we denote the one-step transition probability between states $i$ and $j$ as $\pTrans{i}{j}$, and the stationary probability for the chain to be in state $i$ as $\pi_i$. 

%% file: sysModel.tex
\section{System Model and Preliminaries} \label{sec:sysModel}

We consider a population of \nodes\ terminals (or nodes), each monitoring an underlying stochastic process, and reporting updates to a common receiver over a shared wireless channel. Time is divided in slots of equal duration, and all terminals are assumed to be synchronized to this pattern. The tracked processes are modelled as independent and identically distributed two-state discrete time Markov chains. Without loss of generality, we shall focus on a specific node, and denote the corresponding Markov process, taking values in $\{0,1\}$, as \mcn, $n\in \mathbb N$. State changes can take place at the beginning of each slot, and the stationary distribution of the chain is
\begin{align}
    \statZ = \frac{\qOZ}{\qZO+\qOZ}, \quad \statO = \frac{\qZO}{\qZO+\qOZ}.
\end{align}
Furthermore, we denote by \avgQ\ the average probability for the process to perform a transition, i.e. \mbox{$\avgQ= \statZ \qZO + \statO \qOZ$}, 
 and characterize asymmetric processes ($\qZO\neq\qOZ$) by the ratio $\eta = \qZO/\qOZ$.

Nodes share the channel towards the receiver following a random access policy based on slotted ALOHA \cite{Abramson77:PacketBroadcasting} \emph{without feedback}, independently deciding at each slot whether to transmit the current state of their source. More details will be provided in \secr\ref{sec:policies}. To capture the behavior of the multiple access links we resort to the well-established collision channel model. Accordingly, no data can be retrieved at the receiver over a slot in which two or more terminals transmit, whereas a packet received without collision is  decoded.

The receiver maintains over time an estimate \estn\ of the state of the monitored process, updating it whenever an update on the tracked value is obtained from the corresponding terminal ($\estn = \mcn$), and keeping the previous estimate otherwise ($\est_{n} = \est_{n-1}$). In this context, it is useful to introduce the estimation error at time $n\in\mathbb N$, defined as
\begin{align}
    \errn := \lvert \mcn - \estn \rvert
\end{align}
and taking values in $\{0,1\}$.
We denote for convenience as $\{\treset_i\}$ the sequence of time instants at which the error is reset, i.e., \errn\ transitions from $1$ to $0$, and as $\{\trise_i\}$ the one capturing the instants at which \errn\ moves from $0$ to $1$ (see \figr\ref{fig:timeline}). Note that, for the system under study, the former event may take place in two cases: i) when an update containing the current condition of the source is received; or ii) when the source transitions to the state estimated at the receiver, regardless of whether a message notifying the change is delivered. 

\subsection{Performance metrics}
\begin{figure}[t]
    \centering
    \includegraphics[width=\columnwidth]{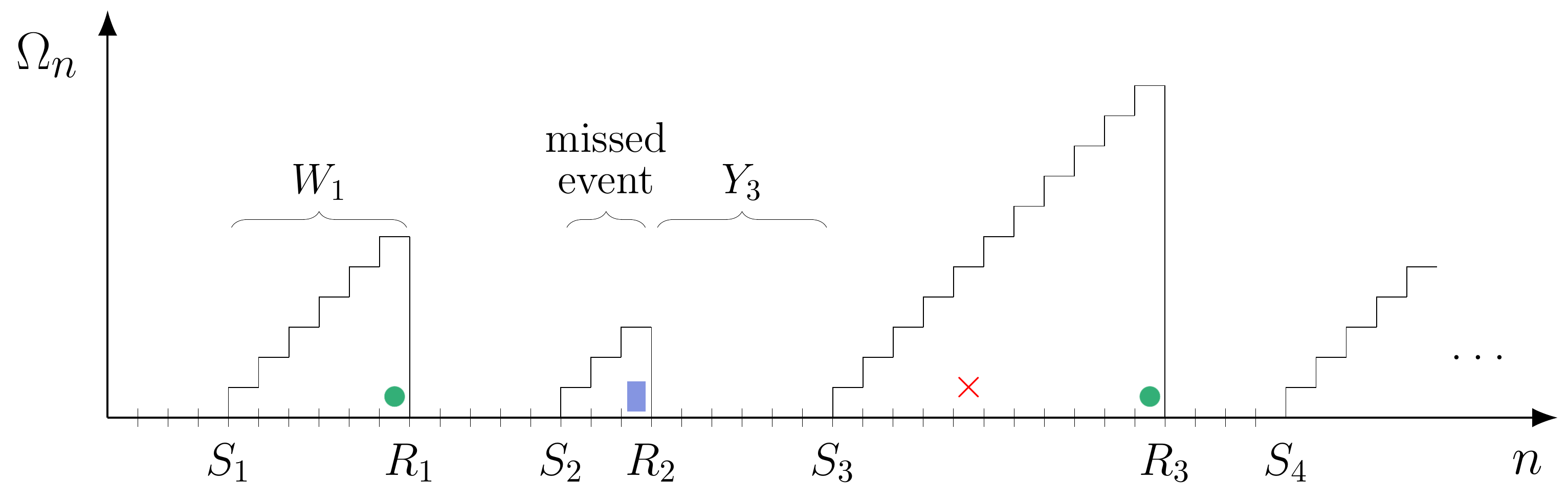}
    \vspace{-1.5em}
    \caption{Example evolution of the AoII over time. A green circle (\mycircle{darkGreen!40}) denotes a slot in which an update is successfully delivered, resetting the metric, whereas a red cross (\textcolor{red}{$\times$}) indicate a failed transmission. Finally, a blue rectangle  (\squareA{oxygenblue!70}) refers to a slot in which the metric is reset due to a non-notified  state change.}
    \label{fig:timeline}
\end{figure}

To gauge the ability of the receiver to maintain an accurate perception of the source state, we resort to the \emph{\acl{AoII}} (AoII) \cite{Ephremides19_AoII}.  The instantaneous value of the metric at time index $n$ is defined as
\begin{align}
    \aoii_n = \errn \cdot \left( n - \trise[n] \right)
\end{align}
where $\trise[n] = \max\{\trise_i \,\vert\, \trise_i \leq n\}$ is the last time at which the receiver started having an erroneous estimate. The definition leads to the evolution exemplified in \figr\ref{fig:timeline}, where no penalty is undergone whenever a correct knowledge is available (\mbox{$\errn=0$}), while a linear loss proportional to the time spent in error is experienced otherwise. The stochastic process $\aoii_n$ can easily be shown to be ergodic, and we will focus in the remainder on its average value
\begin{align}
    \Aoii = \expOp[\aoii_n] = \lim_{\ell\to\infty} \frac{1}{\ell} \sum\nolimits_{i=1}^\ell \aoii_i. 
    \label{eq:aoiiDef}
\end{align}

To complement our study, we consider two additional performance indicators. On the one hand, we are interested in characterizing the average duration of periods during which the receiver has an erroneous estimate of a tracked process. The metric can be particularly relevant for many IoT systems, and its minimization may be especially important whenever actuation based on wrong knowledge is critical. Leaning on the introduced notation, the duration of the $i$-th error period is captured as $\errper_i = \treset_i - \trise_i$, and we will refer to the expected value of the process in stationary conditions as \emph{average error duration}, $\expOp[\errper]$. In the remainder of our analysis, it will also be useful to resort to the r.v. $\corrper_i = \trise_{i+1} - \treset_i$, which describes the duration of a period with correct knowledge ($\errn=0$).

On the other hand, we focus our attention on the possibility of a \emph{missed detection}. The event occurs when a monitored source transitions to a state and leaves it without the receiver noticing. This can be particularly detrimental, possibly leading to the overlook of a critical situation, e.g., requiring proper alarms or counter-actions. The probability of such an event will be denoted as \pmiss. 
 
\begin{figure*}[!b]
    \hrule
    \begin{align}
       \mathtt z \cdot \pi_{00} &= \qOZ (\pchange\qZO + \pstale(1-\qZO)) (\qOZ + (1-\qOZ) \pstale\psucc)
       \qquad \quad 
       \mathtt z \cdot \pi_{01} = \qZO\qOZ (\pchange\qOZ + \pstale(1-\qOZ)) (1 -\pchange\psucc) \\
       \mathtt z \cdot \pi_{11} &= \qZO (\pchange\qOZ + \pstale(1-\qOZ)) (\qZO + (1-\qZO) \pstale\psucc)
       \qquad \quad 
       \mathtt z \cdot \pi_{10} = \qZO\qOZ (\pchange\qZO + \pstale(1-\qZO)) (1 -\pchange\psucc)
       \label{eq:stationary_dist}
    \end{align}
    \begin{align}
       \mathtt z = (\qZO + \qOZ) \{ \pchange\qOZ\qZO(2-\pchange\psucc) + \pstale [ \,(\qZO+\qOZ)(1-\pstale\psucc) + \pstale\psucc - \qZO\qOZ(2-\pstale\psucc) \,]\, 
       \}
    \end{align}
\end{figure*}

\subsection{Channel access strategies} \label{sec:policies}
In the remainder, we consider three variations of the basic ALOHA contention. The simplest case is represented by a \emph{random strategy}, which foresees each node attempt delivery of a message at every slot with probability \prand. This policy completely decouples channel access from the evolution of the monitored source, and can capture well the behavior encountered in may practical IoT systems, e.g., aiming at maximizing throughput or minimizing age of information. 

On the other hand, we also tackle a \emph{reactive strategy}, where a terminal transmits with probability one only when a change of the source state occurs, and remains silent otherwise, aiming to avoid redundant messages.

Finally, we study a \emph{hybrid strategy}, allowing a node to access the channel with probability \pchange\ in case of a state transition, and with probability \pstale\ otherwise.  We remark that the former approaches can be seen as special cases of the hybrid solution, setting $\pchange=\pstale=\prand$ for the random case, and $\pstale=0$, $\pchange=1$ for the reactive one. Leaning on this, the channel load for the access policies, i.e., the average number of transmitted packets per slot, is given by $\load = \avgAct\nodes$, where \avgAct\ is the average probability for a node to attempt a transmission over a slot, and is readily computed as
\begin{align}
    \avgAct = \avgQ \pchange + (1-\avgQ) \pstale.
    \label{eq:avgAct}
\end{align}
Accordingly, we characterize the probability for a transmitted packet to be retrieved at the receiver, as
\begin{align}
    \psucc = (1-\avgAct)^{\nodes-1} \stackrel{(a)}{\simeq} e^{-\load}
    \label{eq:psucc}
\end{align}
where the approximation $(a)$ is very tight for the large values of \nodes\ typical of massive IoT systems. The expression quantifies the probability that a transmitted packet does not undergo a collision, and is exact for i.i.d. behavior of nodes across slots. From this standpoint we remark that the assumption always holds in presence of symmetric sources ($\qZO=\qOZ$, $\eta=1$). For asymmetric transitions ($\eta\neq 1$), instead, the number of sources in a given state may determine the level of channel contention over a slot when the reactive or hybrid policies are employed. Therefore, the equality in \eqref{eq:psucc} holds in this case for the random strategy, and is an approximation otherwise, whose tightness will be discussed in \secr\ref{sec:results}. Finally, we denote the throughput, i.e., the average number of decoded packets per slot, as \tru, and resort to the expression $\tru \simeq \load e^{-\load}$. 

%% file: analysis.tex
\section{Performance Analysis} \label{sec:analysis}

We start our analysis by providing the following result.
\begin{lemma}
For the system model under study, the average AoII is given by
\begin{align}
    \Aoii = \frac{\expOp\left[ \errper^2 \right] + \expOp\left[\errper \right]}{2\left(\expOp\left[ \errper \right] + \expOp\left[ \corrper \right] \right)}.
    \label{eq:aoiiFormula}
\end{align}
\end{lemma}
\begin{IEEEproof}
The derivation follows simple geometric arguments. In particular, we focus without loss of generality on an observation interval of duration $n$, ending with the reset of the AoII process. Accordingly, $n$ can be expressed as the sum of the $k$ intervals of duration $\corrper_i$ and $\errper_i$ that fall into it. Resorting to this, \eqref{eq:aoiiDef} can be written as
\begin{align}
    \Aoii &= \lim_{k\to\infty} \frac{k}{\sum_{i=1}^k (\errper_i + \corrper_i)} \, \frac{1}{k}\sum\nolimits_{i=1}^{k} \sum\nolimits_{\ell=0}^{\errper_i} \ell
\end{align}
where the rightmost summations express the aggregate AoII over the whole period, recalling that the metric is zero over all the intervals $\corrper_i$ and grows linearly over the intervals $\errper_i$. Observing that $\sum\nolimits_{\ell=0}^{\errper_i} \ell = (\errper_i + \errper_i^2)/2$, the result follows by the ergodicity of the involved processes.
\end{IEEEproof}

\begin{figure}
    \centering
    \includegraphics[width=.9\columnwidth]{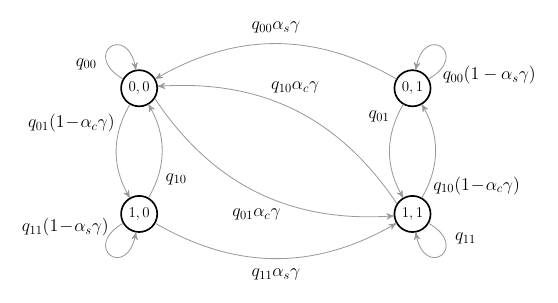}
    \vspace{-1em}
    \caption{Markov chain tracking the joint evolution $(\mcn,\estn)$ of the source state and of the estimate at the receiver.}
    \vspace{-1em}
    \label{fig:markovJoint}
\end{figure}

The reported formulation highlights how \Aoii\ can be obtained from the moments of the r.v.s \errper\ and \corrper. To this aim, consider the Markov chain $(\mcn,\estn)$, jointly tracking the state of the source of interest and the estimate available at the receiver. The transition probabilities for this four-state process can be readily obtained for the general hybrid strategy, and are reported in \figr\ref{fig:markovJoint}. Similarly, the stationary distribution $\pi_{ij}$, $i,j\in\{0,1\}$ follows by directly solving the balance equations, and is reported at the bottom of this page, leaning on the auxiliary quantity $\mathtt z$ for compactness. 


\begin{figure*}[!t]
    \begin{align} 
    \expOp[\errper] \!=\! \frac{\mathtt{c^\prime}}{1\!-\!\qZZ(1\!-\!\pstale\psucc)} + \frac{1-\mathtt{c}^\prime}{1\!-\!\qOO(1\!-\!\pstale\psucc)}, \qquad
    \expOp[\errper^2] \!=\! \frac{\mathtt c^\prime (1\!+\!\qZZ(1\!-\!\pstale\psucc))}{(1\!-\!\qZZ(1\!-\!\pstale\psucc))^2} + \frac{(1-\mathtt c^\prime)(1\!+\!\qOO(1\!-\!\pstale\psucc))}{(1\!-\!\qOO(1\!-\!\pstale\psucc))^2} 
    \label{eq:moments}
\end{align}
\hrule
\end{figure*}

Consider now \errper, i.e., the duration of a period over which the receiver has a wrong estimate. Leaning on the joint process $(\mcn,\estn)$, we observe that one such interval can only be experienced when the chain enters $(0,1)$ or $(1,0)$, and its duration corresponds to the permanence time in the state. Let us focus on the first case, and denote by $\errper_{01}$ the conditional r.v. describing the corresponding value of \errper. Observing that the receiver reverts to a correct knowledge at each slot if state $(0,1)$ is left, $\errper_{01}$ follows a geometric distribution with parameter $1-\qZZ(1-\pstale\psucc)$. Applying a similar reasoning, when the wrong estimate period corresponds to a visit to $(1,0)$ for $(\mcn,\estn)$, the conditional value of \errper\ is characterized by the r.v. $\errper_{10} \sim \text{Geo}(1-\qOO(1-\pstale\psucc))$. In turn, the probability of entering an error period in $(\mcn,\estn) = (0,1)$ is given by $\pi_{11}\qOZ(1-\pchange\psucc)$, whereas a visit to $(1,0)$ has probability $\pi_{00}\qOZ(1-\pchange\psucc)$. Accordingly, the conditional probability of having an error period in $(0,1)$, denoted as $\mathtt{c}^{\prime}$ is
\begin{align}
    \mathtt{c}^\prime = \frac{\pi_{11}\qOZ}{\pi_{11}\qOZ + \pi_{00}\qZO}
\end{align}
and, by the law of total probability, the PMF of \errper\ follows as
$ p(\errperval) = \mathtt{c}^\prime p(\errperval_{01}) + (1-\mathtt{c}^\prime) p(\errperval_{10})$. Finally, the sought moments can be obtained from the geometric distribution of $\errper_{01}$ and $\errper_{10}$, as reported in \eqref{eq:moments} at the top of this page.


The same approach can be used to derive the distribution of \corrper, observing that the conditional probability of terminating a period of correct knowledge when starting from $(0,0)$ and $(1,1)$ is \mbox{$1-\qZZ-\qZO\pchange\psucc$} and \mbox{$1-\qOO-\qOZ\pchange\psucc$}, respectively. Applying the law of total probability and computing the expectation readily leads to
\begin{align}
    \expOp[\corrper] = \frac{\mathtt c^{\prime\prime}}{1-\qZZ-\qZO\pchange\psucc} + \frac{1-\mathtt c^{\prime\prime}}{1-\qOO-\qOZ\pchange\psucc}
\end{align}
where $\mathtt c^{\prime\prime}$ is the probability of having a period with correct knowledge taking place in state $(0,0)$ and evaluates to\footnote{The expression follows by noting, from the balance equations of the Markov chain, that the probability of entering state $(0,0)$ from other states is $\pi_{00} - \pi_{00}\qZZ$. A similar reasoning holds for entry in state $(1,1)$.}
\begin{align}
    \mathtt c^{\prime\prime} = \frac{\pi_{00}(1-\qZZ)}{\pi_{00}(1-\qZZ) + \pi_{11}(1-\qOO)}.
\end{align}
Plugging the statistical moments of \errper\ and \corrper\ into \eqref{eq:aoiiFormula} allows to characterize in closed form the average AoII for any values of the source transition probabilities as well as for all the access strategies being tackled. Consider for example the case of \emph{symmetric sources} ($\qZO=\qOZ=\avgQ$). In such setting, it is easy to verify that $\pi_{00}=\pi_{11}$ and $\pi_{01}=\pi_{10}$, so that, after basic manipulations we obtain the compact formulation
\begin{align}
    \Aoii = \frac{\avgQ\nodes^2(1-\pchange\psucc)\vphantom{\frac{A}{B}}}{[\,\avgQ\nodes(1-\pchange\psucc) + \load\psucc\,] \,[\,2\,\avgQ\nodes(1-\pchange\psucc) + \load\psucc\,]\vphantom{\frac{A}{B}}}.
    \label{eq:aoiiSymmetric}
\end{align}

To conclude, we derive the probability of missed detection. Without loss of generality, let us focus on the specific source state $1$, e.g., assuming it is associated to a critical condition. In such setting, we are interested in the probability that the receiver is not informed of a visit to the state, given that it has taken place. This occurs when two conditions are met: i) the transition to $1$ is not immediately notified to the receiver; and ii) no update is delivered for the whole interval of geometric duration over which the source remains in state $1$. Accordingly,
\begin{align}
    \pmiss &= (1-\pchange\psucc) \sum\nolimits_{\ell=0}^{\infty} (1-\pstale\psucc)^\ell(1-\qOZ)^\ell \qOZ\\
    & = \frac{\qOZ(1-\pchange\psucc)}{\qOZ + \pstale\psucc(1-\qOZ)}.
    \label{eq:pmiss}
\end{align}

%% file: results.tex
\section{Numerical Results and Discussion} \label{sec:results}

\begin{figure}
    \centering
    \includegraphics[width=.9\columnwidth]{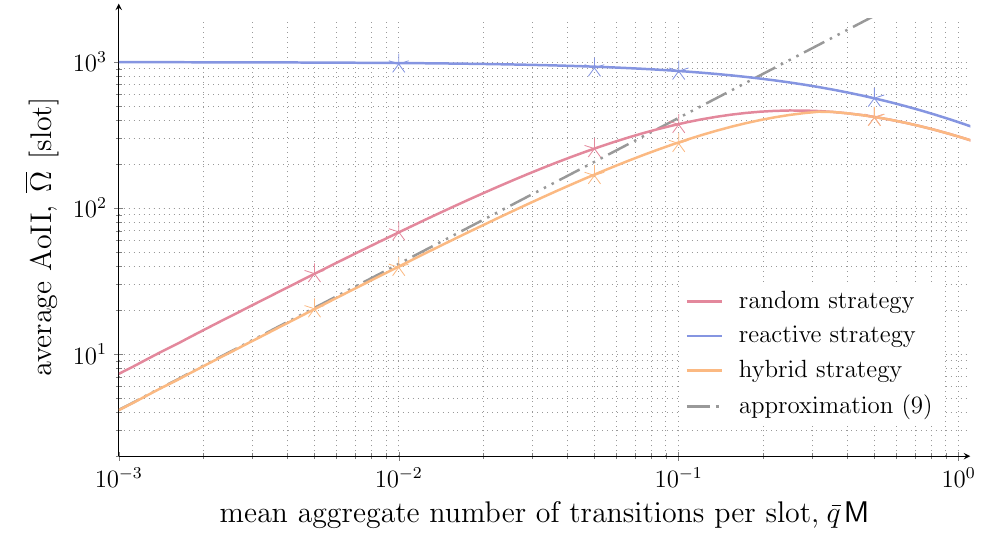}
    \caption{Average AoII vs average aggregate number of state transitions, symmetric sources. For random and hybrid strategies, transmission probabilities are tuned for any \avgQ\ to minimize \Aoii. Markers denote simulation results.}
    \vspace{-1em}
    \label{fig:aoiiVsQ}
\end{figure}

To gauge the performance of the considered access schemes  we first consider the average AoII \Aoii\ in the symmetric case. The metric behavior is reported in \figr\ref{fig:aoiiVsQ} against the average number of aggregate state changes per slot in the system, $\avgQ\nodes$, providing a measure of how dynamic the sources are. In the plot, solid lines denote analytical results obtained via \eqref{eq:aoiiSymmetric}, whereas markers the outcome of numerical simulations.
All shown results are obtained for $\nodes=1000$ nodes.

Let us start by studying the leftmost region of the plot, characterized by low values of $\avgQ$. Note that such conditions correspond to sporadic node activity, and are those typical of many IoT applications. Focus first on the reactive strategy (blue curve). In this case, $\pchange=1$, $\pstale=0$, with channel load $\load = \avgQ \nodes$. When $\avgQ \ll 1$, simple manipulations on \eqref{eq:aoiiSymmetric} leaning for \psucc\ on the approximation  $e^{-\load} \simeq 1 - \avgQ\nodes$ lead to $\Aoii \simeq \nodes$, as pinpointed by the figure. The sole transmission in case of state change does not offer thus good performance, leaving the receiver with an erroneous estimate for large fractions of time. Indeed, following a reactive strategy, whenever a packet is lost due to collision, the receiver remains in error until at least the next state transition for the source, as no other notification will be sent by the the corresponding node in the meantime. As \avgQ\ decreases, such periods become longer, with a detrimental impact on \Aoii\ which is proportional to the number of contending terminals.

Secondly, consider the random strategy, for which $\pchange=\pstale=\alpha$, and $\load = \alpha\nodes$, irrespective of \avgQ. To characterize the behavior of this approach, it is convenient to express the average AoII in terms of the throughput \tru, rearranging \eqref{eq:aoiiSymmetric} as
\begin{align}
    \Aoii = \frac{\avgQ \,\nodes (\nodes-\tru)}{[\,\avgQ\nodes + \tru(1-\avgQ)\,] \,[\,2\,\avgQ\nodes + \tru(2-\avgQ)\vphantom{A^B}\,]}.
    \label{eq:aoiiRandom}
\end{align}
It is then easy to infer that \Aoii\ is minimized by maximizing the throughput, i.e., by setting the access probability as $\alpha = 1/\nodes$ to attain $\tru=e^{-1}$.
 The reported results (red line) are accordingly obtained with this parameter choice. Moreover, for  $\avgQ \ll 1$, \eqref{eq:aoiiRandom} provides the useful approximation $\Aoii \simeq \avgQ\,\nodes^2/\tru^2$, showing how the average AoII scales as $7.39 \cdot \avgQ \nodes^2$ for sporadic nodes activity. The improvement with respect to the reactive solution is evident in \figr\ref{fig:aoiiVsQ}. From this standpoint, although operating the shared channel at peak throughput entails a higher packet loss probability ($\psucc\simeq e^{-1} = 0.36$) compared to solely transmitting in the presence of a state change ($\psucc \simeq e^{-\avgQ\nodes}$), the possibility to notify the current value of the tracked source more often pays off by reducing the duration of error periods and thus overall lowers AoII.

Finally, we turn our attention the the hybrid approach. In this case, two degrees of freedom are available in tuning the access policy, and the natural question of how to set $(\pchange,\pstale)$ arises. In this regard, a joint minimization of \eqref{eq:aoiiSymmetric} with respect to the transmission probabilities does not lead to a compact mathematical formulation, and is omitted for the sake of compactness, noting that the problem can easily be solved numerically. More interestingly, we summarize some key take-aways. First, observing that for low values of $\avgQ\nodes$ both factors in the denominator of \eqref{eq:aoiiSymmetric} are driven by $\load\psucc$, the expression can be suitably approximated as
\begin{align}
    \Aoii \simeq \frac{\avgQ\,\nodes^2 (1-\pchange\psucc)}{(\load e^{-\load})^2}.
    \label{eq:aoiiHybrid}
\end{align}
From the expression we infer that \pchange\ shall be set to $1$. Indeed, any pair $(\pchange,\pstale)$ leading to the same channel load \load\ but using $\pchange<1$ would result in a higher average AoII. Leaning on this, the load $\load^*$ at which the system shall be operated to minimize \Aoii\ can be obtained by simply nulling the derivative of \eqref{eq:aoiiHybrid} with respect to \load. The optimal value follows as the unique strictly positive solution of the transcendental equation
\begin{align}
    2(\load-1) e^{\load} = \load-2
\end{align}
obtaining $\load^* \simeq 0.644$ [pkt/slot]. Recalling \eqref{eq:avgAct}, the optimal probability of transmitting in the absence of a state change can immediately be derived, so that overall we have \mbox{$(\pchange^*,\pstale^*) = (1, \load^*/[\,\nodes(1-\avgQ)\,])$} for $\avgQ \ll 1$. 

In this perspective, two remarks are in order. On the one hand, we observe that  minimizing AoII requires operation at a channel load strictly smaller than $1$, entailing a cost in terms of throughput. Notably, thus, a tradeoff between the efficiency of a slotted ALOHA channel and AoII emerges. On the other hand, plugging the optimal access probabilities in \eqref{eq:aoiiHybrid} leads to the scaling law $\Aoii \simeq 4.51 \cdot \avgQ \nodes^2$ \--- reported in \figr\ref{fig:aoiiVsQ} by the dash-dotted line. Comparing this to the trends discussed earlier, we conclude that a hybrid strategy almost halves for $\avgQ\nodes \ll 1$ the experienced AoII with respect to the random approach, enabling  a reduction of \Aoii\ of a factor $0.56$.

As the probability of state transition for a source increases, the numerical optimization of $(\pchange,\pstale)$ reveals how the former parameter shall initially be kept to $1$. By doing so, the channel load progressively increases, as nodes become active more frequently by reporting changes in the monitored processes. This solution eventually becomes detrimental, leading to channel congestion. When this happens, a sudden change occurs, and the optimal implementation of a hybrid solution collapses onto the random approach by having $\pchange^*=\pstale^*=1/\nodes$. This is clearly highlighted in \figr\ref{fig:aoiiVsQ}, where the gain of the hybrid scheme (yellow curve) progressively vanishes, converging for sufficiently large values of $\avgQ\nodes$ to the random one (red curve).

Finally, \figr\ref{fig:aoiiVsQ} provides another relevant insight. We remark in fact that, in the rightmost region of the plot, \Aoii\ decreases, in spite of more frequent source state changes which would require more resources to be correctly notified. This counterintuitive trend stems from the very definition of the AoII. Indeed, an erroneous knowledge at the receiver can be corrected not only with the delivery of an update, but also in case the corresponding source transitions to the state which is currently estimated, even in the absence of any communications. Both cases have the same effect, inducing a reset of $\aoii_n$. For large values of $\avgQ\nodes$, the latter situation prevails, lowering the average AoII.\footnote{Note that, for two-state sources, $\Aoii\to1/2$  as $\avgQ\to 1$.} In this perspective medium access plays a little role, and the policies offer similar performance.

\begin{figure}
    \centering
    \includegraphics[width=.9\columnwidth]{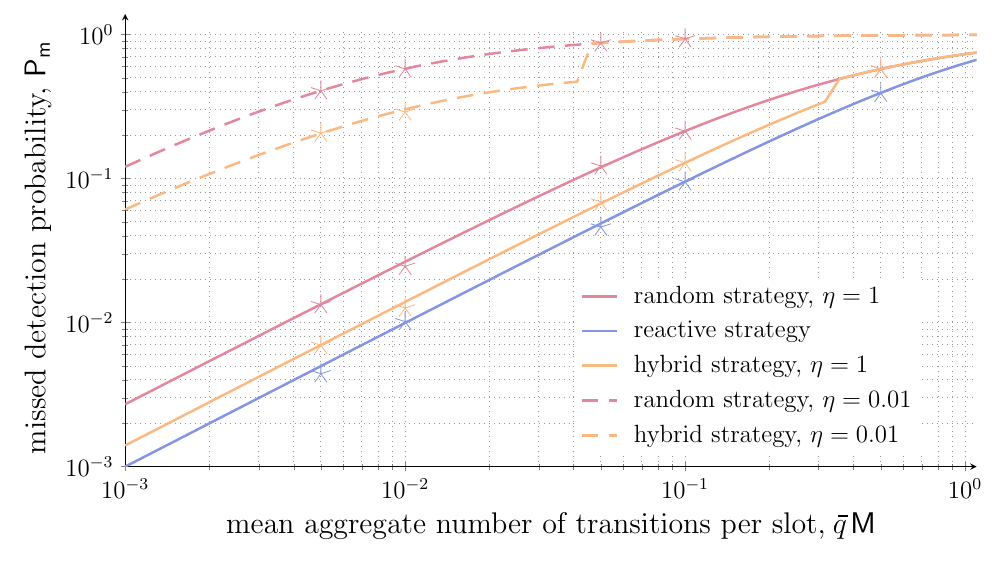}
    \caption{Missed detection probability vs average number of state transitions in the network, $\avgQ\nodes$. Solid lines denote symmetric sources, whereas dashed lines asymmetric sources $\eta=0.01$. Markers denote simulation results.}
    \vspace{-1em}
    \label{fig:pMissVsQ}
\end{figure}

To further elaborate on this, we report \figr\ref{fig:pMissVsQ}, which shows the missed detection probability computed in \eqref{eq:pmiss} when varying $\avgQ\nodes$. For both the random and the hybrid strategy, the access probabilities were set to minimize AoII.  Let us first consider solid lines in \figr\ref{fig:pMissVsQ}, referring to the symmetric case ($\eta=1$). As expected, \pmiss\ grows with $\avgQ\nodes$, due to two factors. On the one hand, quicker transitions between source states are more difficult to be tracked at the receiver, as a shorter time to report them is available for the nodes. On the other hand, the increased level of source activity poses more challenges to the medium access policies in terms of update delivery. For the reactive approach, this translates  into an increased channel contention (and thus higher packet loss probability). In turn, the fixed channel load attained by the random strategy comes at the expense of a higher chance of not reporting an event at all. Overall, it appears clear that the rightmost region of the plot is of scarce practical relevance for most IoT systems, corresponding to a too high probability of missing source transitions.
Interestingly, the figure reveals that a reactive approach performs better in terms of missed detection, pinpointing a fundamental trade-off with the average AoII. Along this line, we observe that the hybrid policy improves over the random one, thanks to immediate notification of source state changes, yet loses with respect to the reactive scheme due to the increased packet loss rate induced by setting $\pstale>0$.\footnote{As discussed, the quick change in the hybrid policy behavior denotes the switching point after which \pchange\ is set to $1/\nodes$, reverting to a random scheme.}

Fig.~\ref{fig:pMissVsQ} also reports the trends obtained in presence of asymmetric sources (dashed lines), for $\eta=0.01$. For the reactive case, no performance change is noted. Indeed, as predicted by \eqref{eq:pmiss}, the missed detection probability for the policy evaluates to $\pmiss=1-e^{-\avgQ\nodes}$, only depending on the average source change rate \avgQ. For the random and the hybrid solutions, instead, a significantly higher missed detection rate is undergone. This stems from the shorter time spent in state $1$ for the asymmetric case ($\qZO = 0.01\, \qOZ$), which renders the detection of transitions among states harder. The result is particularly interesting, as settings in which visits to a critical state are sporadic and possibly short may be of practical relevance in many anomaly detection or remote monitoring applications. Incidentally, we note that an excellent match is obtained between simulation results (markers) and analytical trends (dashed line) also for the asymmetric hybrid case, proving the validity the approximation of \psucc\ in \secr\ref{sec:sysModel}.

\begin{figure}
    \centering
    \includegraphics[width=.9\columnwidth]{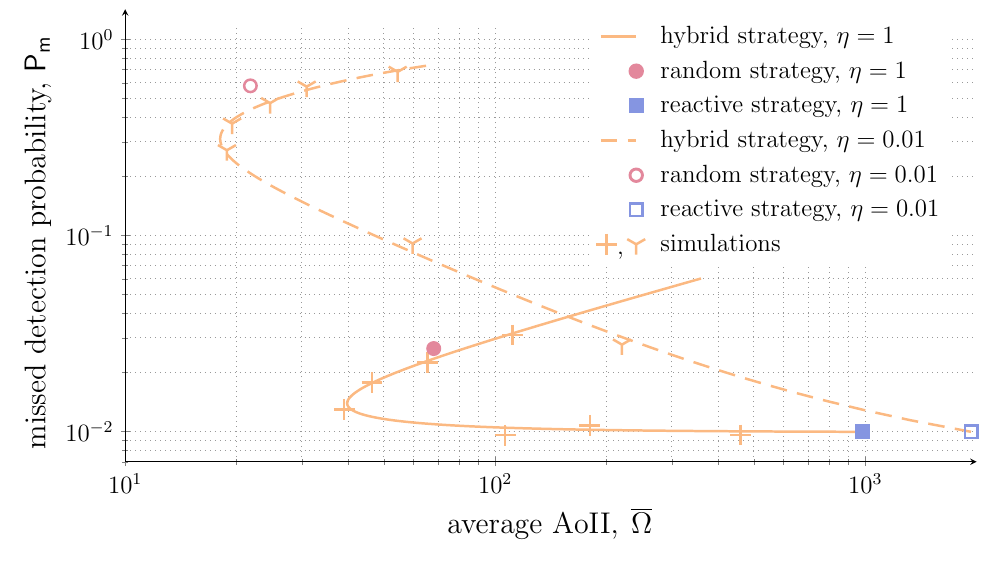}
    \vspace{-1em}
    \caption{Average AoII vs missed detection probability, for $\avgQ\nodes=10^{-2}$. Star markers denote simulation results.}
    \vspace{-1em}
    \label{fig:contourPm}
\end{figure}

The discussion carried out so far has pinpointed a trade-off between AoII and missed detection. This aspect is further explored in \figr\ref{fig:contourPm}, reporting \Aoii\ vs \pmiss\ for \mbox{$\avgQ\nodes = 10^{-2}$} for both symmetric and asymmetric ($\eta=0.01$) sources. For the reactive (square markers) and random (circle markers) strategies, a single point is reported, showing the $(\Aoii,\pmiss)$ pair achieved with the corresponding access probabilities. Conversely, a complete curve is available for the hybrid scheme, obtained by varying $\pstale$ in $[0,3/\nodes]$ while keeping $\pchange = 1$ (clearly, for $\pstale=0$ the curve coincides with the reactive scheme marker). A fundamental design choice emerges, as low values of missed detection probability can be attained at the cost of an increased AoII, especially for asymmetric sources. In this respect, the considered hybrid scheme proves particularly useful, allowing to strike the desired balance between the two performance metrics by tuning \pstale. The presented framework offers thus a useful tool, providing simple analytical expressions to suitably select the access protocol parameters.

\begin{figure}
    \centering
    \includegraphics[width=.86\columnwidth]{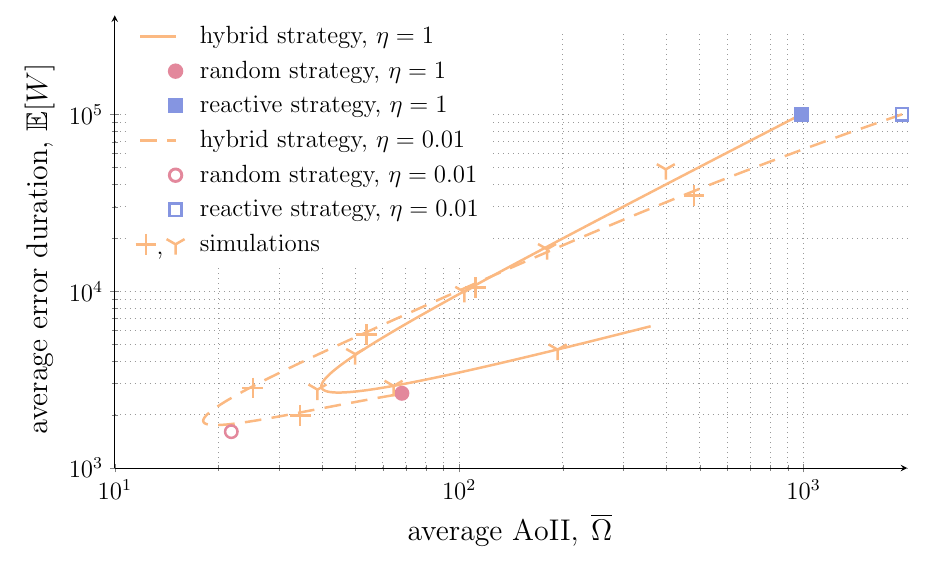}
    \vspace{-1em}
    \caption{Average AoII vs average duration of an error period, for $\avgQ\nodes=10^{-2}$. Star markers denote simulation results.}
    \vspace{-1em}
    \label{fig:contourOut}
\end{figure}

To conclude our study we present in \figr\ref{fig:contourOut} the joint behavior of the average AoII and of the average error duration, $\expOp[\errper]$ for $\avgQ\nodes=10^{-2}$. Also in this case, the curves for the hybrid strategy were obtained by varying \pstale\ with \mbox{$\pchange=1$}, while markers indicate the performance of the random and reactive solutions. As expected, the latter scheme performs the worst, since transmissions only in the presence of a state change trigger long periods of outdated knowledge at the receiver in case of packet loss. In turn,  the effect can be significantly reduced by allowing nodes to send updates also in the absence of a change. Notably, a tradeoff emerges also between \Aoii\ and $\expOp[\errper]$ for the hybrid policy, with the optimal values of the two being attained for different configurations of the access parameters. Although less pronounced than what observed in \figr\ref{fig:contourPm}, the result further stresses the importance of properly  tuning the access strategy in remote source monitoring systems to optimize the metric of most interest.

%% file: conclusions.tex
\section{Conclusions} \label{sec:conclusions}

We tackled the problem of remote source monitoring in IoT networks over random access channels. Specifically, we consider a population of devices that share a slotted ALOHA channel \emph{without feedback} to a common receiver, sending updates on the value of independent two-state Markov chains. At the receiver, a simple estimator which stores the last obtained value from each node is implemented. The systems aims at maintaining an up-to-date knowledge on the monitored processes, and performance is measured in terms of three metrics: average age of incorrect information, probability of missed detection (intended as the probability for the receiver to completely miss a source transition to a critical state), and average duration of the periods spent with an erroneous estimate. Different variations of the channel access policies are studied: a random approach which completely decouples transmissions from the source behavior, a reactive one which foresees a node to transmit only in case of a source transition, and a hybrid solution, which allows a terminal to transmit with different probabilities depending on whether a change in the source state has occurred or not. In all cases we derive analytical closed form expressions for the performance indicators, deriving the optimal transmission probabilities and highlight some fundamental trends. We pinpoint a trade-off between AoII and missed detection probability, showing that the channel access approach shall be carefully selected based on the metric of interest for the application being served.